\documentclass[twocolumn,showpacs]{revtex4}
\usepackage{graphicx}
\usepackage{dcolumn}
\usepackage{bm}
\usepackage{amsmath}
\usepackage{amsfonts}
\usepackage{amssymb}

\providecommand{\U}[1]{\protect\rule{.1in}{.1in}}

\begin{document}

\title{The quantum walk temperature}
\author{Alejandro Romanelli}
\altaffiliation{alejo@fing.edu.uy}
\affiliation{Instituto de F\'{\i}sica, Facultad de Ingenier\'{\i}a\\
Universidad de la Rep\'ublica\\
C.C. 30, C.P. 11300, Montevideo, Uruguay}
\date{\today }

\begin{abstract}
A thermodynamic theory is developed to describe the behavior of the
entanglement between the coin and position degrees of freedom of the
quantum walk on the line. This theory shows that, in spite of the
unitary evolution, a steady state is established after a Markovian
transient stage. This study suggests that if a quantum dynamics is
developed in a composite Hilbert space (\emph{i.e.} the tensor
product of several sub-spaces) then the behavior of an operator that
only belongs to one of the sub-spaces may camouflage the unitary
character of the global evolution.
\end{abstract}

\pacs{03.67.-a, 03.65.Ud, 02.50.Ga}
\maketitle

\section{Introduction}

The concept of isolated system plays a fundamental role in the
formulation of the quantum mechanics. This concept is an
idealization that was constructed as an aid to understand some
phenomena displayed by real systems which may be regarded as
approximately isolated. However, since about 50 years ago, the study
of quantum decoherence has acquired a central position in the
formulation of the quantum mechanics. In fact, concepts such as
thermodynamic equilibrium seem impossible to coordinate with the
idea of isolated system because the quantum state for such a system
follows a unitary evolution and it cannot reach a final equilibrium state at $%
t\rightarrow\infty $.

In this context we ask ourselves if it is possible to introduce the
concept of temperature for an isolated quantum system which evolves
in a composite Hilbert space. In this paper the system known as the
quantum walk on the line \cite{QW} has been chosen as a model to
answer this question . The quantum walk (QW) is a natural
generalization of the classical random walk in the frame of quantum
computation and quantum
information processing and it is receiving permanent attention \cite%
{childs,Linden,Alejo3}. It has the property to spread over the line
linearly in time as characterized by the standard deviation $\sigma
(t)\sim t$, while its classical analog spreads out as $\sigma
(t)\sim t^{1/2}$. This property, as well as quantum parallelism and
quantum entanglement, could be used to increase the efficiency of
quantum algorithms \cite{Shenvi,Childs et}. Recently we have been
investigating \cite{alejo2010,alejo2011,armando2011} the asymptotic
behavior of the QW on the line, focusing on the probability
distribution of chirality independently of position. We showed that
this distribution has a stationary long-time limit that depends on
the initial conditions. This result is unexpected in the context of
the unitary evolution of the QW because such a behavior is usually
associated to a Markovian process. In this paper  we further explore
the behavior of the chirality distribution and define a
thermodynamic equilibrium between the degrees of freedom of position
and chirality.  This equilibrium allows to introduce a temperature
concept for this unitary closed system. We obtain a master equation
with a time-dependent population rate, that describes the transient
behavior of  the reduced density operator of the QW towards
thermodynamic equilibrium. The QW's reduced density operator shows
an surprising behavior: Its behavior looks diffusive but however the
global evolution of the system is unitary.

The paper is organized as follows. In the next section the standard
QW model is developed, in the third section the entanglement
temperature is defined. Then in the fourth and fifth sections the
entanglement temperature is obtained for localized initial
conditions and for distributed initial conditions respectively. In
the six section the transient behavior towards thermal equilibrium
is studied, and finally in the last section some conclusions are
drafted.

\section{QW on the line}
 The composite Hilbert space of the QW is the tensor product
$\mathcal{H}_{s}\otimes \mathcal{H}_{c}$ where $\mathcal{H}_{s}$ is
the Hilbert space associated to the motion on the line and
$\mathcal{H}_{c}$ is the chirality (or coin) Hilbert space. In this
composite space the walker moves, at discrete time steps $t\in
\mathbb{N}$, along a one-dimensional lattice of sites $k\in
\mathbb{Z}$. The  direction of motion depends on the state of the
chirality, with the eigenstates $R$ and $L$. The wave vector can be
expressed as the spinor
\begin{equation}
|\Psi (t)\rangle =\sum\limits_{k=-\infty }^{\infty }\left[
\begin{array}{c}
a_{k}(t) \\
b_{k}(t)%
\end{array}%
\right] |k\rangle ,  \label{spinor}
\end{equation}%
where the upper (lower) component is associated to the left (right)
chirality.

Then $P_{kL}(t)=\left\vert a_{k}(t)\right\vert ^{2}$ and $%
P_{kR}(t)=\left\vert b_{k}(t)\right\vert ^{2}$ denote the
probability of finding the walker at $\left( k,t\right) $ and the
coin in state $R$ and $L$, respectively. The probability of finding the walker at $%
\left( k,t\right) $ is
\begin{equation}
P(k,t)=\left\langle \Psi _{k,t}\right. \left\vert \Psi _{k,t}\right\rangle
=\left\vert a_{k}(t)\right\vert ^{2}+\left\vert b_{k}(t)\right\vert ^{2},
\label{prob}
\end{equation}%
and $\sum_{k}P(k,t)=1$.

The QW is ruled by a unitary map whose standard form is
\cite{Romanelli09,Alejo2,Alejo1,Alejo4}
\begin{align}
a_{k}(t+1)& =a_{k+1}(t)\,\cos \theta \,+b_{k+1}(t)\,\sin \theta ,\,  \notag
\\
b_{k}(t+1)& =a_{k-1}(t)\,\sin \theta \,-b_{k-1}(t)\,\cos \theta .
\label{mapa}
\end{align}%
where $\theta \in \left[ 0,\pi /2\right] $ is a parameter defining
the bias of the coin toss ($\theta =\frac{\pi }{4}$ for an unbiased
or Hadamard coin). The global left and right chirality probabilities
is defined as
\begin{align}
P_{L}(t)& \equiv \sum_{k=-\infty }^{\infty }P_{kL}(t)=\sum_{k=-\infty
}^{\infty }\left\vert a_{k}(t)\right\vert ^{2},\,  \notag \\
P_{R}(t)& \equiv \sum_{k=-\infty }^{\infty }P_{kR}(t)=\sum_{k=-\infty
}^{\infty }\left\vert b_{k}(t)\right\vert ^{2},  \label{chirality}
\end{align}%
with $P_{R}(t)+P_{L}(t)=1$. The global chirality distribution (GCD)
is defined as the distribution formed by the couple $\left[
\begin{array}{c}
P_{L}(t) \\
P_{R}(t)%
\end{array}%
\right] $. It is shown in Ref. \cite{alejo2010} that the GCD
satisfies the following map
\begin{align}
{\left[
\begin{array}{c}
P_{L}(t+1) \\
P_{R}(t+1)%
\end{array}%
\right] }& ={\left(
\begin{array}{cc}
\cos ^{2}\theta  & \sin ^{2}\theta  \\
\sin ^{2}\theta  & \cos ^{2}\theta
\end{array}%
\right) }\left[
\begin{array}{c}
P_{L}(t) \\
P_{R}(t)%
\end{array}%
\right]   \notag \\
& +\mathrm{Re}\left[ Q(t)\right] \sin {2}\theta \left[
\begin{array}{c}
1 \\
-1%
\end{array}%
\right] ,  \label{master}
\end{align}%
where
\begin{equation}
Q(t)\equiv \sum_{k=-\infty }^{\infty }a_{k}(t)b_{k}^{\ast }(t).  \label{qdet}
\end{equation}%
The two dimensional matrix in Eq.(\ref{master}) can be interpreted
as a transition probability matrix for a classical two dimensional
random walk as it satisfies the necessary requirements, namely, all
its elements are positive and the sum over the elements of any
column or row is equal to one. On the other hand, it is clear that
$Q(t)$ accounts for the interferences. When $Q(t)$ vanishes the
behavior of the GCD can be described as a classical Markovian
process. However in the generic case $Q(t)$ together with $P_{L}(t)$
and $P_{R}(t)$ are time depend functions that  have long-time
limiting values \cite{alejo2010} which are determined by the initial
conditions of Eq.(\ref{mapa}).  Eq.(\ref{master}) can be solved in
this limit.  We define
\begin{align}
\Pi _{L}& \equiv
\begin{array}{c}
\lim \text{ }P_{L}(t) \\
t\rightarrow \infty~~~~
\end{array}%
,\,  \notag \\
\Pi _{R}& \equiv
\begin{array}{c}
\lim \text{ }P_{R}(t) \\
t\rightarrow \infty~~~~
\end{array}%
,\,  \notag \\
Q_{0}& \equiv
\begin{array}{c}
\lim \text{ }Q(t) \\
t\rightarrow \infty~~
\end{array}%
,\,  \label{asym}
\end{align}%
and then  we obtain the asymptotic stationary solution for the GCD
as
\begin{equation}
{\left[
\begin{array}{c}
\Pi _{L} \\
\Pi _{R}%
\end{array}%
\right] }=\frac{1}{2}\left[
\begin{array}{c}
1+2\mathrm{Re}(Q_{0})/\tan \theta  \\
1-2\mathrm{Re}(Q_{0})/\tan \theta
\end{array}%
\right] .  \label{estacio}
\end{equation}
This interesting result for the QW shows that the long-time
probability to find the system with left or right chirality has a
limit. Therefore, although the dynamical evolution of the QW is
unitary, the evolution of its GCD has an asymptotic limit
characteristic of a diffusive behavior. This situation is further
surprising if we compare our case with the case of the QW on finite
graphs \cite{Aharonov} where it is shown that there is no converge
to any stationary distribution.

\section{Entanglement and temperature}

The concept of entanglement is an important element in the
development of quantum communication, quantum cryptography and
quantum computation. In this context several authors have
investigated the relation between asymptotic
entanglement and the initial conditions of the QW \cite%
{Carneiro,abal,Annabestani,Omar,Pathak,Petulante,Venegas,Endrejat,Ellinas1,Ellinas2,Maloyer}%
. Other authors \cite{Venegas1,Chandrashekar} have proposed to use
the QW as a tool for quantum algorithm development and as  an
entanglement generator potentially useful to test quantum hardware.

The unitary evolution of the QW generates entanglement between the
coin and position degrees of freedom. To characterize this
entanglement  we start with the von Neumann entropy which is quantum
analogue of the Gibbs entropy
\begin{equation}
S_{N}(\rho )=-\mathrm{tr}(\rho \log \rho ).  \label{uno}
\end{equation}%
where $\rho =|\Psi (t)\rangle \langle \Psi (t)|$ is the density
matrix of the quantum system. Due to the unitary dynamics of the QW
the system remains in a pure state and this entropy vanishes.
However for these pure states the entanglement between the chirality
and the position can be quantified by
the associated von Neumann entropy for the reduced density operator \cite%
{Carneiro,abal} that defines the entropy of entanglement
\begin{equation}
S(\rho )=-\mathrm{tr}(\rho _{c}\log \rho _{c}),  \label{dos}
\end{equation}
where
\begin{equation}
\rho _{c}=\mathrm{tr}(\rho ),  \label{dos}
\end{equation}
and the partial trace is taken over the positions.
Using the wave function Eq.(\ref{spinor}) and its normalization
properties we obtain the reduced density operator \cite{Carneiro}
\begin{equation}
\rho_{c} =\left(
\begin{array}{cc}
P_{L}(t) & Q(t) \\
Q(t)^{\ast } & P_{R}(t)%
\end{array}%
\right) .  \label{rho}
\end{equation}
The density operator $\rho_{c}$ has the eigenvalues
\begin{equation}
\lambda _{\pm}=\frac{1}{2}\left[ 1\pm \sqrt{1-4\left(
P_L(t)\,P_R(t)-\left\vert Q(t)\right\vert ^{2}\right) }\right].  \label{lam}
\end{equation}
Then the entanglement entropy Eq.(\ref{dos}) is expressed through these
two eigenvalues as
\begin{equation}
S(\rho)=-\lambda_{+}\log \lambda_{+}-\lambda_{-}\log \lambda_{-}.
\label{ttres}
\end{equation}%
In the asymptotic regime $t\rightarrow\infty$ the eigenvalues go to
a stationary limit, $\lambda _{\pm}\rightarrow\Lambda _{\pm}$ and
\begin{equation}
\Lambda _{\pm}=\frac{1}{2}\left[ 1\pm \sqrt{1-4\left(\Pi_L\,\Pi_R-\left\vert
Q_0\right\vert ^{2}\right) }\right].  \label{lam0}
\end{equation}
Further from Eq.(\ref{estacio}) follows the relation
\begin{equation}
\Pi_L\,\Pi_R= \frac{1}{4}-{\left(\frac{\mathrm{Re}(Q_{0})}{\tan \theta}%
\right)}^{2},  \label{piqu}
\end{equation}
which is substituted in Eq.(\ref{lam0}), and then the asymptotic
eigenvalues are expressed as
\begin{equation}
\Lambda _{\pm}=\frac{1}{2}\pm \sqrt{\chi},  \label{lam1}
\end{equation}
with
\begin{equation}
\chi\equiv \left\vert Q_0\right\vert ^{2}+{\left({\mathrm{Re}(Q_{0})}/{\tan
\theta}\right)}^{2}.  \label{defi}
\end{equation}
Note that the values of the interference term $Q_{0}$ are constrained to
satisfy the condition
\begin{equation}
0<\Lambda _{+}\Lambda _{-}<1,  \label{condition0}
\end{equation}
and then
\begin{equation}
0<\chi<\frac{1}{4}.  \label{condition}
\end{equation}
\begin{figure}[th]
\begin{center}
\includegraphics[scale=0.38]{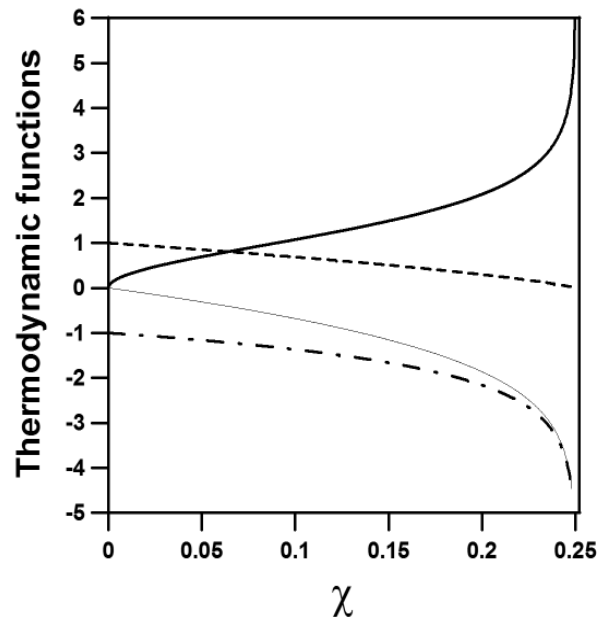}
\end{center}
\caption{Dimensionless thermodynamic function normalized by $\log(2)$ as a
function of the dimensionless parameter $\protect\chi$. From the top to the
bottom they are: in thick line $\protect\beta\protect\epsilon$, in dashed
line the entropy $S_0$, in thin line the energy $\protect\beta U$ and in
dashed-dot line the Helmholtz free energy $\protect\beta A$.}
\label{f1}
\end{figure}
The entanglement entropy has an asymptotic limit too
\begin{equation}
S_0=-\Lambda _{+}\log \Lambda _{+}-\Lambda _{-}\log \Lambda _{-},
\label{s0}
\end{equation}
that only depends on the initial conditions through the interference term $%
Q_{0}$. Therefore we are led to consider that after some transient
time the QW achieves a thermodynamic equilibrium between the
position and chirality degrees of freedom. In order to make a fuller
description of this equilibrium it is necessary to connect the
eigenvalues of $\rho_c$ with its associated Hamiltonian operator
$H_c$.  To obtain this connection we shall use the  quantum Brownian
motion model of Ref.\cite{Kubo}. We considered the system associated
with the chirality degrees of freedom and characterized by the
density matrix $\rho_c$ in thermal contact (entanglement) with the
bath system associated with the position degrees of freedom, the
lattice. In this context $\rho_c$ satisfies the equation
\begin{equation}
\frac{\partial\rho_{c}}{\partial
t}=\frac{1}{i\hbar}[H_c,\rho_{c}]+\Gamma \rho_{c}, \label{scho}
\end{equation}
where $[H_c,\rho_{c}]$ is the commutator and $\Gamma \rho_{c}$
represents the Brownian motion of $\rho_c$ induced by the noise
(fluctuating forces) exerted on $\rho_c$ by the lattice (position
degrees of freedom). In the equilibrium (stationary) situation we
must have ${\partial\rho_{c}}/{\partial t}= 0$ and $\Gamma
\rho_{c}=0$ \cite{Kubo}, that is
\begin{equation}
[H_c,\rho_{c}]=0. \label{scho2}
\end{equation}
Therefore, in the asymptotic regime, the density operator $\rho_{c}$
must be an explicit function of the Hamiltonian operator which must
be time independent. Now we call $\{\Phi_{+},\Phi_{-}\}$ the
eigenfunctions of the density matrix, and then the operators $H_{c}$
and $\rho_{c}$ are both
diagonal in this basis. Therefore, the eigenvalues $\Lambda _{+}$ and $%
\Lambda _{-}$ depend of the corresponding eigenvalues of $H_c$. We
take these eigenvalues to be $\{-\epsilon,\epsilon\}$ without any
loss of generality, they represent the possible values of the
entanglement energy. This
interpretation agrees with the fact that $\Lambda _{+}$ and $%
\Lambda _{-}$ are the probabilities that the system is in the
eigenstate $\Phi_{+}$ or $\Phi_{-}$. The precise dependence between
$\Lambda _{\pm}$ and $\pm \epsilon$ is determined by the type of
ensemble we construct.
The main proposal of this paper is that this equilibrium corresponds
to a quantum canonical ensemble. Therefore we propose that
\begin{equation}
\Lambda _{\pm}\equiv\frac{e^{\pm\beta\epsilon}}{e^{\beta\epsilon}+e^{-\beta%
\epsilon}},  \label{lam2}
\end{equation}
which defines the entanglement temperature $T\equiv1/\beta$. Then
the probability that the state chosen at random from the ensemble $\{\Phi_{+}
$, $\Phi_{-}$\}, possesses an energy $\epsilon$ is determined by the
Boltzmann factor $e^{-\beta\epsilon}$. Let us call $\widetilde{\rho_{c}}$ the
diagonal expression of the density operator $\rho_{c}$, then
\begin{equation}
\widetilde{\rho_{c}}=\left(
\begin{array}{cc}
\Lambda _{+} & 0 \\
0 & \Lambda _{-}%
\end{array}%
\right) =\frac{1}{e^{\beta\epsilon}+e^{-\beta\epsilon}}\left(
\begin{array}{cc}
e^{\beta\epsilon} & 0 \\
0 & e^{-\beta\epsilon}%
\end{array}%
\right) .  \label{rhoni}
\end{equation}
This operator is formally the same density operator that corresponds
to an electron with possesses an intrinsic spin and a magnetic
moment in a external magnetic field \cite{pathria}.

Starting from Eq.(\ref{rhoni}) it is possible to build the thermodynamics for
the QW entanglement. The partition function of the
system is then given by
\begin{equation}
\mathcal{Z}=e^{\beta\epsilon}+e^{-\beta\epsilon}=2\cosh({\beta\epsilon}).
\label{betaA}
\end{equation}
Accordingly, and also using Eqs.(\ref{lam0},\,\ref{lam2}), the
temperature is given by
\begin{equation}
T=2\epsilon/\ln\left(\frac{1+2\sqrt{\chi}}{1-2\sqrt{\chi}}\right),
\label{betae}
\end{equation}
the Helmholtz free energy by
\begin{equation}
A=-\frac{1}{\beta}\ln[2\cosh({\beta\epsilon})]=\frac{T}{2}\ln\left(\frac{1}{4}%
-\chi\right),  \label{betaA}
\end{equation}
the internal energy by
\begin{equation}
U=-\epsilon\tanh(\beta\epsilon)=-2\epsilon\sqrt{\chi}, \label{betaU}
\end{equation}
and finally the entropy by
\begin{equation}
S_0=\beta U-\beta A,  \label{s02}
\end{equation}
where this last thermodynamic definition for the entropy of course
agrees with the previous Shannon expression in Eq.(\ref{s0}). To
finished this section in Fig.~\ref{f1} we present the dependence of
these thermodynamic magnitudes with the interference parameter
$\chi$.

\section{Localized initial conditions}

\begin{figure}[th]
\begin{center}
\includegraphics[scale=0.38]{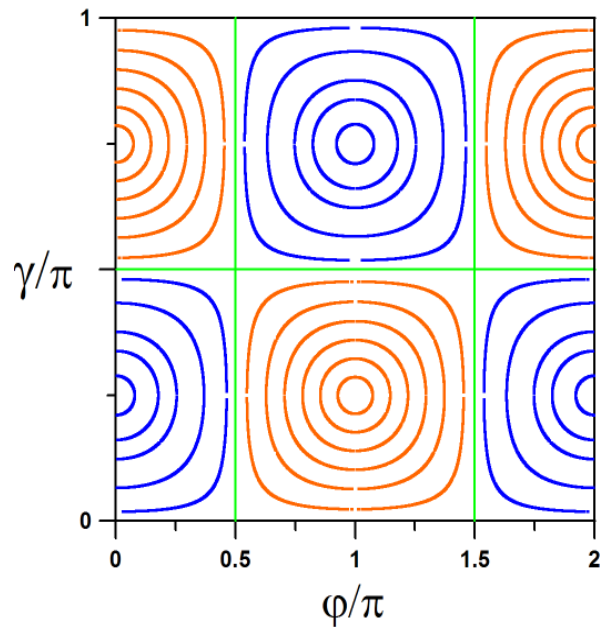}
\end{center}
\caption{Isothermal curves as functions of the dimensionless angles $%
\protect\gamma$ and $\protect\varphi$. Due to the rotation symmetry
in the angle $\varphi$, only four zones are distinguished: two
"cold" and two "hot". The "hot" zones (orange on line) have six
isotherms; from inside to outside their temperatures are: $%
T/T_{0}=6.5,~3.2,~2.2,~1.6,~1.3$ and $1.1$. The "cold" zones (blue
on line) have five isotherms; from outside to inside their
temperatures are: $T/T_{0}=0.9,~0.8,~0.7,~0.68$ and $0.66$. The
straight (green) lines corresponded to $T/T_{0}=1$, see
Eq.(\ref{varphygamma}).} \label{f2}
\end{figure}
\begin{figure}[th]
\begin{center}
\includegraphics[scale=0.34]{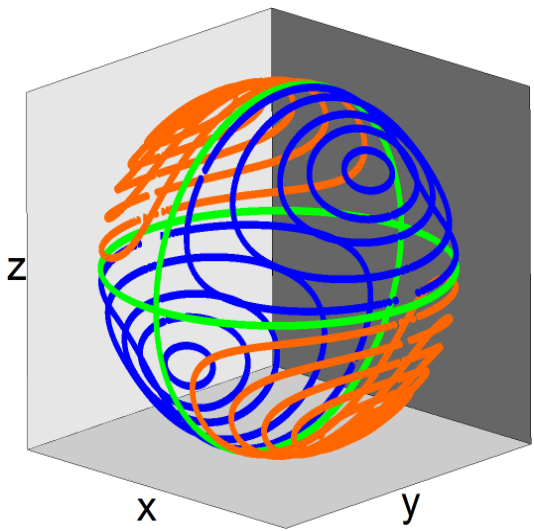}
\end{center}
\caption{The isothermal curves of Fig.~\ref{f2} shown on the Bloch
sphere. The QW initial chirality determines the entanglement
temperature, see Eqs.(\ref{betae},\ref{psi0},\ref{chi2}).}
\label{f3}
\end{figure}
As seen in the previous section the thermodynamics of entanglement
only depends on the interference term $Q_{0}$ which in turn only
depends on the initial conditions, as shown in \cite{alejo2010}.

In order to investigate this dependence on the initial conditions of the system
we consider first the localized case. The
initial state of the walker is assumed to be sharply localized at the
origin with arbitrary chirality, thus
\begin{equation}
|\Psi (0)\rangle =\left(
\begin{array}{c}
\cos ({\gamma}/{2}) \\
\exp i\varphi \text{ }\sin ({\gamma}/{2})%
\end{array}%
\right) |0\rangle ,  \label{psi0}
\end{equation}%
where $\gamma \in \left[ 0,\pi \right]$ and $\varphi \in \left[ 0,2\pi %
\right] $ define a point on the unit three-dimensional Bloch sphere.
The expression of $Q_{0}$
was obtained in Ref. \cite{abal}, fixing
the bias of the coin toss $\theta =\pi /4$,
following the method developed by Nayak and
Vishwanath \cite{nayak}
\begin{equation}
Q_{0}=\frac{1}{2}(1-\frac{1}{\sqrt{2}})\left[ \cos \gamma \text{ }+\sin
\gamma \text{ }(\cos \varphi +i\sqrt{2}\sin \varphi )\right] .
\label{q0entengl}
\end{equation}
Using this result in Eq.(\ref{defi}) the dependence of $\chi$ with the
initial conditions is given by
\begin{equation}
\chi=\chi_{0}\left(1+ \cos \varphi \sin 2\gamma \right) ,  \label{chi2}
\end{equation}
where $\chi_{0}={3}/{4}-1/\sqrt{2}$.

It is useful to define a characteristic temperature (in units of
$\epsilon$)
\begin{equation}
T_{0}=2/ \left[\ln\left(\frac{1+2\sqrt{\chi_{0}}}{1-2\sqrt{\chi_{0}}}\right)%
\right],  \label{t0}
\end{equation}
in order to express any other temperature as a proportionality with $%
T_0=1/\beta_0$. Then from Eq.(\ref{betae}) we obtain an expression
for $\beta$ as a function of the angles $\gamma$ and $\varphi$
\begin{equation}
\cos\varphi \sin
2\gamma=\left(\frac{\tanh\beta}{\tanh\beta_0}\right)^2-1.
\label{varphygamma}
\end{equation}
Figures ~\ref{f2} and \ref{f3} show the level curves (isotherms) for
the entanglement temperature  as a function of the QW initial
position. In Fig.~\ref{f2} the initial position is defined through
the angles $\gamma$ and $\varphi$ and in Fig.~\ref{f3} it is defined
through the position on the Bloch sphere (see Eq.(\ref{psi0})). Both
figures show four regions, two of them corresponding to temperatures
$T>T_0$ (orange color on line) and the other two to temperatures
$T<T_0$ (blue color on line). The longest isotherms (green color on
line) correspond to the temperature $T=T_{0}$ and their initial
conditions are $\gamma=0,~\pi/2,~\pi$ and $\varphi=\pi/2,~3\pi/2$.

\section{Distributed initial conditions}

\begin{figure}[th]
\begin{center}
\includegraphics[scale=0.38]{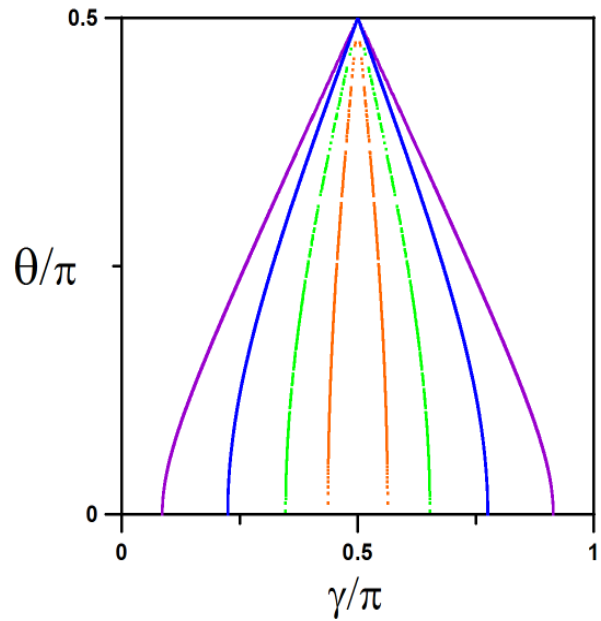}
\end{center}
\caption{Isothermal curves as functions of the dimensionless angles $%
\protect\gamma$ (initial conditions) and $\protect\theta$ (bias of
the coin). Four curves (with different colors on line) are
presented; each curve
has two branches placed symmetrically on both sides of $\protect\gamma=%
\protect\pi/2$, where $T=\infty$. The values of $T$ are given by
Eqs.(\ref{betae2}). From left to right the values of $T$ are, in
units of $\protect\epsilon$: 0.5 (purple), 1. (blue), 2. (green) and
5. (orange). The diagram has two discontinuities in
$\theta=\protect\pi/2$ and in $\protect\gamma=\protect\pi/2$ (see
Eqs.(\protect \ref{betae2},\protect\ref{re2})).} \label{f4}
\end{figure}
In previous works \cite{Eugenio,alejo2010} we have studied the QW
with extended initial conditions. Now the entanglement temperature
is studied in such a case. The following extended Gaussian
distributions is proposed:
\begin{equation}
a_{k}^{0}\equiv{\left[ \frac{1}{\sigma_0\sqrt{2\pi }}\exp \left( -\frac{k^{2}%
}{2\sigma_0^{2}}\right) \right]}^{\frac{1}{2}}\cos({\gamma}/{2}) \text{,}
\label{aes}
\end{equation}%
\begin{equation}
b_{k}^{0}\equiv e^{i \varphi }{\left[ \frac{1}{\sigma_0\sqrt{2\pi }}\exp
\left( -\frac{k^{2}}{2\sigma_0^{2}}\right) \right]}^{\frac{1}{2}}\, \sin({%
\gamma}/{2})\text{,}  \label{bes}
\end{equation}%
where $\sigma_0$ is the initial standard deviation, $\gamma \in \left[ 0,\pi %
\right]$ determines the initial proportion of the left and right
chirality and $\varphi\in \left[ 0,2\pi \right] $ is a global phase.
Using these initial conditions, Eqs.(\ref{aes}, \ref{bes}), the
asymptotic value of $Q(t)$, see Eqs.(\ref{qdet},\ref{asym}), was
obtained \cite{alejo2010} as
\begin{equation}
Q_0=\frac{1}{2}\cos\gamma\,\tan\theta\,,  \label{q1}
\end{equation}
with the restrictions
\begin{equation}
\sigma_0\gg1,  \label{sig1}
\end{equation}
and
\begin{equation}
\cos\varphi=\frac{\tan\theta}{\tan\gamma}.  \label{phi2}
\end{equation}
Replacing Eq.(\ref{q1}) in Eq.(\ref{defi}) we obtain
\begin{equation}
\chi= \left( \frac{\cos\gamma}{2\cos\theta}\right)^{2},  \label{chides}
\end{equation}
and then using Eq.(\ref{betae}) we have
\begin{equation}
\beta\epsilon=\frac{1}{2}\ln\left(\frac{\left|\cos\theta\right|
+\left|\cos\gamma\right|}{\left|\cos\theta\right|-\left|\cos\gamma\right|}%
\right),  \label{betae2}
\end{equation}
where taking into account Eqs.(\ref{condition},\,\ref{chides}) the
initial condition satisfies the constraint
\begin{equation}
{\left|\cos\gamma\right|}<{\left|\cos\theta\right|}.  \label{re2}
\end{equation}

The functions $Q_0$, $\chi$ and $\beta$ vanish for $\gamma=\pi/2$ (
see Eqs.(\ref{q1},\,\ref{chides},\,\ref{betae2})) and simultaneously
the entanglement entropy Eq.(\ref{ttres}) has its maximum value $S_0
= 1$. This maximum value is achieved when the entanglement
temperature is $T=\infty$. Under these conditions the system behaves
as a classical Markov process
\cite{alejo2010}. On the other hand, the initial conditions $\gamma$ and $%
\varphi$ are not independent (see Eq.(\ref{phi2})) and for each value of $%
\gamma$ there is only one value of $T$, then for fixed $\theta$, it
is not possible to have isotherms as functions of $\gamma$ and
$\varphi$. Instead the entanglement temperature depends on $\theta$
and $\gamma$ from Eq.(\ref{betae2}), \emph{i.e-} the choice of the
bias of the coin toss $\theta$ or of the initial proportion of the
chirality $\gamma$ could lead to the same entanglement temperature.
Fig.~\ref{f4} shows the isotherms as functions of $\gamma$ and
$\theta$.

\section{Transient behavior}

\begin{figure}[th]
\begin{center}
\includegraphics[scale=0.38]{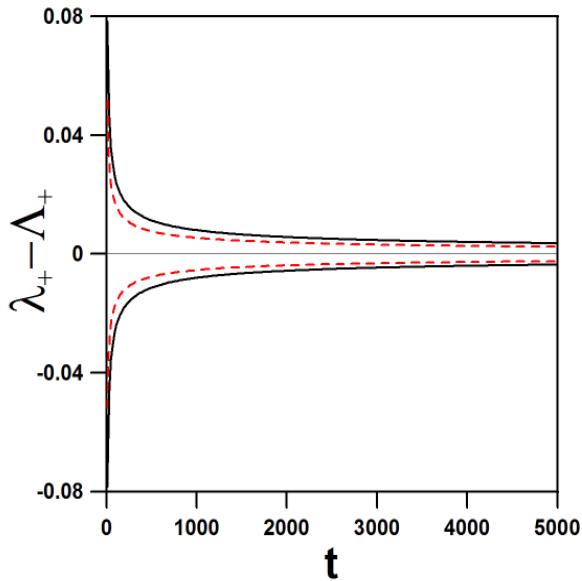}
\end{center}
\caption{Envelope of the probability $\protect\lambda _{+}-\Lambda
_{+}$ as a function of the dimensionless time $t$ for two different
initial conditions. Each initial condition is established by the
couple $(\varphi, \gamma)$ in Eq. (\protect\ref{psi0}). Their values
are $(\protect\pi /8, \protect\pi /4)$ for the full black line and
$(\protect\pi/4, \protect\pi /3) $ for the dashed red line. In both
cases, the temperature is $T=0.79~T_{0}$.} \label{f5}
\end{figure}
\begin{figure}[th]
\begin{center}
\includegraphics[scale=0.36]{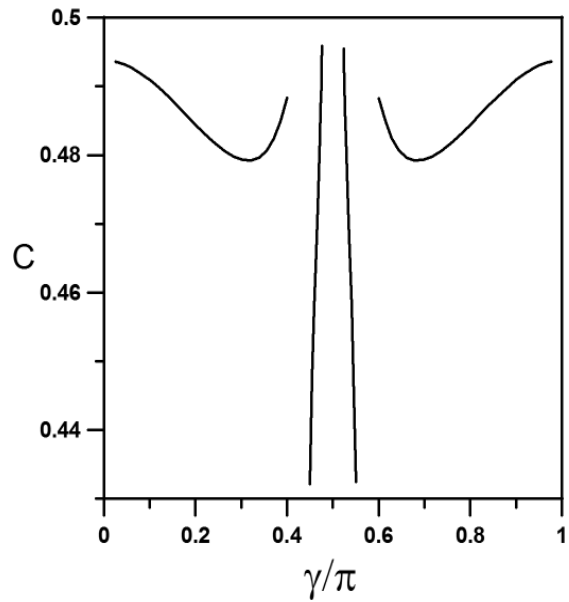}
\end{center}
\caption{The power law exponent as a function of the dimensionless angle $%
\protect\gamma $. The initial conditions $(\varphi, \gamma)$
correspond to the isotherms $T=1.1~T_{0}$.} \label{f6}
\end{figure}

In the QW a stationary entanglement is established between the
chirality and position degrees of freedom after a transient time.
This fact allowed us to introduce the concept of entanglement
temperature. The transient behavior of the system is studied using
the original map Eq.(\ref{mapa}) in a numerical code with initial
conditions given by Eq.(\ref{psi0}). These numerical calculations
are summarized in Figs. ~\ref{f5} and ~\ref{f6}. Fig.~\ref{f5}
presents the difference between the transient ($\lambda _{+}$) and
the stationary ($\Lambda _{+}$) eigenvalues of the density matrix as
a function of time (see Eqs.(\ref{lam},\,\ref{lam0})). The figure
only presents the envelope of the curves because the real
eigenvalues dynamics is very intricate; it presents quick
oscillations with high density of paths where it is only possible
distinguish its global contour. However, the average evolution of
the system is determined by the envelope dynamics.
Each envelope has two branches placed symmetrically on both sides of $%
\lambda _{+}-\Lambda _{+}=0$. Two pair of curves are presented in
full and dashed lines with color black and red respectively. In both
cases, the envelopes decay for $t\rightarrow \infty $ as a power law
$1/t^{c}$ with $c=0.490$ for the dashed red line and $c=0.486$ for
the full black line. The envelopes of $\lambda _{\pm}$ will be
called $\widetilde{\lambda}_{\pm}$ respectively.

It was numerically verified for several initial conditions given by
Eq.(\ref{psi0}) that the transient behavior of
$\widetilde{\lambda}_{\pm}-{\Lambda}_{\pm}$ can be adjusted by a
power law function of time. Fig.~\ref{f6} shows the power law
exponent $c$ as a function of the initial condition $\gamma $ for
the same temperature. Remember that, for $T$ and $\gamma $ given,
$\varphi $ is determined by Eq.(\ref{varphygamma}). Therefore, the
exponent $c$ has a dependence with the initial conditions, however
this dependence is not determined by the asymptotic temperature
value.

Additionally, the transient behavior of
$\widetilde{\lambda}_{\pm}-{\Lambda}_{\pm}$ was numerically studied,
using initial conditions given by Eqs.(\ref{aes},\,\ref{bes}) with
$\sigma\gg1$. In these cases, the system showed a negligible
transient dynamics, in concordance with the calculation developed in
Ref. \cite{alejo2010}. Therefore, for these initial conditions, the
reduced density matrix is essentially always in thermodynamic
equilibrium.

With the aim to understand the transient behavior of the system we
develop an analytic theory implementing a parallelism between the
reduced density operator Eq.(\ref{rhoni}) and the density operator
of an electron in a external magnetic field. With this picture on
mind we propose the following master equation for the probabilities
${\lambda} _{+}$ and ${\lambda}_{-}$
\begin{align}
\frac{d{\lambda}_{-}}{dt}& ={\lambda}_{+}\,w_{+-}\,-%
{\lambda}_{-}\,w_{-+},\,  \notag \\
\frac{d{\lambda}_{+}}{dt}& ={\lambda}_{-}\,w_{-+}\,-%
{\lambda}_{+}\,w_{+-},  \label{masterfin}
\end{align}
where $w_{+-}$ and $w_{-+}$ are transition probabilities per unit of
time, that can be understood as population rates.
$w_{+-}$ corresponds to the transition ${\lambda}%
_{+}\rightarrowtail{\lambda}_{-}$ and $w_{-+}$ corresponds to the
transition ${\lambda}_{-}\rightarrowtail{\lambda}_{+}$. These rates
are time-dependent functions, and  their behaviors are known in the
limit $t\rightarrow\infty$ when
${d{\lambda}_{\pm}}/{dt}\longrightarrow 0$. In this limit, the
stationary solution of Eq.(\ref{masterfin}) must be the couple
$\Lambda _{-}$ and $\Lambda _{+}$, given by Eq.(\ref{lam0}). Then
the asymptotic values of the population rates satisfy
\begin{equation}
\frac{w_{b}}{w_{a}}= \frac{\Lambda _{-}}{\Lambda _{+}},  \label{rate}
\end{equation}
where $w_{a}$ and $w_{b}$ are defined by
\begin{align}
w_{a}& \equiv
\begin{array}{c}
\lim \text{ }w_{-+} \\
t\rightarrow \infty~~~
\end{array}%
,\,  \label{rates} \\
w_{b}& \equiv
\begin{array}{c}
\lim \text{ }w_{+-} \\
t\rightarrow \infty~~~
\end{array}%
.\,  \label{rates1}
\end{align}%
Eq.(\ref{rate}) expresses a condition of detailed balance which says
that the rate of occurrence for any transition equals the rate for
the inverse transition. Using our knowledge about the transient and
asymptotic behaviors, the following population rates are proposed
\begin{align}
w_{+-}=w_{b}+\xi(t),\,  \label{rate2}\\
w_{-+}=w_{a}-\xi(t),\,  \label{rate3}
\end{align}
where
\begin{equation}
\xi(t)=\frac{K}{t^{c}}[\omega\sin(\omega
t+\delta)+(\frac{c}{t}-w_{a}-w_{b})\cos(\omega t+\delta)],
\label{xi}
\end{equation}
with $c>0$, $K$, $\omega$ and $\delta$ constants. The general
solution of Eq.(\ref{masterfin}) with these population rates is
\begin{align}
{\lambda}_{+}=\Lambda _{+}+\frac{K}{t^{c}}\cos(\omega t+\delta)+d\,
e^{-(w_{a}+w_{b})t},\,  \label{ratefin}\\
{\lambda}_{-}=\Lambda_{-}-\frac{K}{t^{c}}\cos(\omega t+\delta)-d\,
e^{-(w_{a}+w_{b})t},\,  \label{ratefin1}
\end{align}
where $d$ is an additional constant. Note that $e^{-(w_{a}+w_{b})t}%
\rightarrow0$ faster than ${1/t^{c}}$ for $t\rightarrow\infty$, then
Eqs.(\ref{ratefin},\ref{ratefin1}) verify the asymptotic behavior
obtained numerically. All the constants $K$, $c$, $\omega$, $\delta$
and $d$ depend on the initial conditions, however their values
should be compatible with the positive character of the functions
$\lambda_{+}$, $\lambda_{-}$, $w_{+-}$ and $w_{-+}$. In particular,
to describe correctly the numerical results, $K$ takes a finite
value for localized initial conditions and  $K$ takes negligible
value for distributed initial conditions. In summary, the Brownian
motion equation for our reduced density matrix Eq.(\ref{scho}),
takes the form of a master equation Eq.(\ref{masterfin}).

Finally it is important to point out that the asymptotic behavior
found here is similar to the behavior of the simple cellular
automaton known as \emph{sandpile} \cite{btw}. Such behaviors are
characteristic of extended dynamical systems with spatial degrees of
freedom. They naturally evolve to self-organized states with
correlations that decay with a power law.

\section{Conclusion}
The unitary evolution of the QW in a composite Hilbert space is
studied. In particular the entanglement between chirality and
position degrees of freedom is investigated. After a transient time
the system establishes a stationary entanglement between the coin
and the position that allows to develop a thermodynamic theory. The
asymptotic reduced density operator is used to introduce the
entanglement thermodynamic functions in the canonical equilibrium.
These thermodynamic functions characterize the asymptotic
entanglement. It is shown that the QW initial condition determines
the system's temperature, as well as other thermodynamic function. A
map for the isotherms is analytically built for arbitrary localized
initial conditions. Additionally, it is shown that choosing
appropriately the bias of the coin-toss, it is possible to obtain a
predetermined entanglement temperature.

The transient dynamics of the reduced density operator outside the
thermodynamic equilibrium is also studied. We show numerically that
this transient behavior can be adjusted with a power law, whose
exponent depends on the initial conditions. We built a master
equation to describe this behavior where the population rates have a
time dependence. The accuracy of the master equation solution is
numerically verified and it is shown that the reduced density has a
cellular automaton behavior.

The behavior of the reduced density operator looks diffusive but it
has a dependence with the initial conditions, the global evolution
of the system is unitary. Then, if an observer only had information
related with the chirality degrees of freedom, it would be very
difficult for him to recognize the unitary character of the quantum
evolution. In general, from this simple model we can conclude that
if the quantum system dynamics is developed in a composite Hilbert
space, then the behavior of operators that only belong to one
sub-space could camouflage the unitary character of the global
evolution.


I acknowledge stimulating discussions with V\'{\i}ctor Micenmacher, Guzm\'{a}%
n Hern\'{a}ndez, Ra\'{u}l Donangelo and Armando P\'{e}rez and the support
from PEDECIBA and ANII.

\end{document}